\documentclass[twocolumn,
floatfix,showpacs,
amsmath,amssymb,pre,showpacs]{revtex4}

\usepackage{graphicx}
\usepackage{amsmath}
\usepackage{bm}
\usepackage{isolatin1}
\newcommand{\de}{\partial}

\renewcommand{\v}[1]{\boldsymbol{#1}}

\DeclareGraphicsExtensions{.eps}

\begin{document}

\title{Phase transitions of extended-range probabilistic cellular
 	automata with two absorbing states}

\author{Franco Bagnoli}
 \altaffiliation{also INFN, sez. Firenze and CSDC}
 \email{franco.bagnoli@unifi.it}
 \affiliation{Dipartimento di Energetica, Universit\`a di Firenze,\\
 Via S. Marta 3, 50139 Firenze, Italy}
\author{Fabio Franci}
	\email{fabio@dma.unifi.it} 
 \affiliation{Centro Interdipartimentale per lo Studio delle Dinamiche
 Complesse (CSDC), \\Universit\`a di Firenze, \\
 Via Sansone 1,
50019 Sesto Fiorentino, Italy}
\author{Ra\'ul Rechtman}
 \email{rrs@cie.unam.mx}
 \affiliation{Centro de Investigaci\'on en Energ\'\i a, Universidad   
 Nacional Aut\'onoma
 de M\'exico, \\Apdo.\ Postal 34, 62580 Temixco, Mor., Mexico\\
 }

\date{\today}

\begin{abstract}
We study phase transitions in a long-range one-dimensional cellular
automaton with two symmetric absorbing states. It includes and extends
several other models, like the Ising and Domany-Kinzel ones. 
It is characterized by a competing ferromagnetic linear coupling and an
antiferromagnetic nonlinear one. 
Despite its simplicity, this model exhibits an extremely rich phase diagram.
We present numerical results and mean-field approximations. 
\end{abstract}


\maketitle


\section{Introduction}

Simple discrete models are valuable tools for exploring phase
transitions, both in equilibrium and out of equilibrium. A
paradigmatic example of the first kind is the kinetic Ising model
(heat-bath dynamics),  which can be formulated as a Probabilistic 
Cellular Automaton (PCA) (see Appendix and Ref.~\cite{Georges}). 
One of the simplest
examples of a system exhibiting out of equilibrium phase transitions
is the Domany-Kinzel  (DK)  model~\cite{DP}, which is a natural
extension of the directed percolation (DP) problem in the language of
cellular automata. 

Frustrations play a central role in characterizing the phase diagrams of many 
simple models. One can have geometric frustrations, like in a
triangular  antiferromagnetic Ising model, or fluctuating coupling,
like in spin  glasses~\cite{spinglass} or p-spin~\cite{pspin} models.
 For instance, in the p-spin
model, the character  (ferro or anti-ferro) of the interaction varies
widely with the local  magnetization (i.e. even a flip of a single
spin may invert the  interaction). We discuss here a simpler one dimensional
model in which frustrations play a central role.
This model is a variation of the droplet model in which the particles
tend to repel each other when they are dispersed but in which clusters, once
formed, cannot break and only particles near the surface can eventually leave
them. 

The model originated from the schematization of the mechanism of opinion
formation in a homogeneous (no leader) society~\cite{acri02}, in which it was
assumed that 
a coherent local community exerts a very strong social pressure on an
individual's opinion. However, there is the possibility of disagreement
with a weak local majority depending on the individual's education. 
In the case of a diffuse non-conformistic attitude, people tend to act in the
opposite way of the local majority (antiferromagnetic coupling), introducing
frustrations. The presence of absorbing states may cause the
formation of large coherent clusters, whose interactions give way to
interesting patterns.

This model could also be applied to a system of charged,
magnetized metallic particles. These repel each other due to their charge, but
are attracted due to their magnetic coupling. With a proper choice of 
parameters, 
one may have repulsion when particles are scattered. However, 
once a cluster
of particles in contact appears, the mobility of charges may lower the electric
field among this cluster and an external particle, and the interaction may
become attractive. Other examples can be found in aggregation models. 

In what follows, we discuss a one dimensional spin model 
characterized by  two
coupling factors; one that behaves like the Ising (magnetic)
coupling, i.e., for a given spin, grows linearly with the local
field, and a \emph{nonlinear term} that is very low for a weak local
field, but grows quickly when the local field exceeds a certain
threshold.

In one
dimension, no true phase transition appears for finite-range and
finite-strength coupling. On the other hand, models presenting absorbing
states, like the DK one, do exhibit nonequilibrium phase transitions. The
presence of absorbing states may be related to the divergence of some coupling,
that becomes infinite (see Appendix and Ref.~\cite{Georges}). Thus, in our
model, we simply assume a two-level nonlinear coupling: zero for local field
below a given threshold  (corresponding to the parameter $Q$ in the following)
and infinite  otherwise. Thus, in total we have four parameters: the range
$R$, the external magnetic field $H$ and local coupling $J$ like in the Ising 
model, and the threshold $Q$. We denote these models by $RQ$, indicating
the range and the threshold, say $R3Q1$, $R5Q2$, \dots 

The model can also be discussed as a one dimensional probabilistic
cellular automaton. In this guise, it may be  considered an extension
of the DK model, or, more precisely, of the $R3Q1$ model~\cite{Bagnoli_2001}.
This model exhibits a richer
phase diagram than the DK one, and we shall show that for
$R\to\infty$, the $RQ$ model presents novel features, like a
disorder-chaotic transition. 

It is not surprising that frustrations gives rise to disordered
behavior, which may be sometimes considered chaotic from a
microscopic point of view. We investigate this aspect using an
original chaotic observable, corresponding to a finite-distance
Lyapunov exponent.  However, when renormalized (or locally averaged),
a ``disordered'' behavior  simply corresponds to a stable almost-zero
magnetization. In the mean-field approximation, it corresponds to a
fixed point of observables.  In our model also a coherently-chaotic
phase appears, in which large patches of sites oscillate widely,
corresponding to a chaotic map in the mean-field approximation. 

In summary, we observe that the \emph{usual} transitions from 
ordered into active phases becomes much richer,  being preceded by a
transition from  coherently chaotic to simply chaotic, then  to more
irregular states, the appearance of a parameter-insensitive
disordered phase and finally an  ``escape'' to ordered (quiescent)
states with  a discontinuous (first-order) character.  The origin of
this behavior is the competition  between ``linear''
antiferromagnetic and ``nonlinear'' ferromagnetic couplings. 

The paper is organized as follows. In Sec.~\ref{sec:model} we present
the $RQ$ model both as a spin system and a PCA. In
Sec.~\ref{sec:results} we present numerical results of the model
beginning with a review of the  simplest non trivial case $R=3$,
$Q=1$ and Sec.~\ref{sec:approximations} is devoted to a mean field
discussion of the model.  The paper ends with some conclusions.


\section{The model}
\label{sec:model}

Both the kinetic Ising model (see Appendix) and the DK one  are
Boolean, one dimensional, $R=2$ PCA in which the state of a given
cell depends probabilistically on the sum $S$  of the states of its
two nearest neighbors  at the previous time step. That is, there are
three different  conditional probabilities, $\tau(1|S)$, $S=0,1,2$ that the
future state  of the cell is 1 given $S$ neighbors in state 1. In the
DK model $\tau(1|0)=0$ which means that there is an absorbing phase
corresponding to the configuration where the state on every cell is
zero.  The presence of more than one absorbing state can induce
different behaviors and trigger the appearance of universality 
classes different from the usual directed
percolation (DP)  one~\cite{Hinrichsen_1997,Janssen_1981,Grassberger_1982}. 

We denote by $s_i^t\in\{0,1\}$ the state at site $i=0,\dots,L-1$ and
time $t=0,1\dots$. All operations on spatial indices are assumed to
be modulo $L$ (periodic boundary conditions). The range of
interactions is denoted by $R$, that is, $s_i^{t+1}$ depends on 
the states at time $t$ in a neighborhood ${\cal N}_i$ that contains
the $R$ nearest neighbors of site $i$. It is
convenient to introduce also the spin notation: $\sigma_i^t = 2
s_i^t-1$, $\sigma_i^t \in \{-1,1\}$. 

The spin $\sigma_i^t$ ``feels''  the local field $V(m_i^t) = 
H+[J +K(m_i^t)] m_i^t$, where $H$ is an external field, $J$ a coupling
constant and the local magnetization is defined as
\[
 m_i^t=\sum_{j\in{\cal N}_i} \sigma_j^t=2S_i^t-R,
\]
where
\[
 S_i^t=\sum_{j\in{\cal N}_i} s_j^t.
\]
The presence of absorbing states is due to the nonlinear coupling $K$
given by 
\[
  K(m)=K(m(S,R),Q) = \begin{cases} 
    -k & \text{if $S<Q$,}\\
    k & \text{if $S > R-Q$,}\\
    0 & \text{otherwise.}\\
    \end{cases}
\] 
with $k$ a constant.
This term is relevant only if it is dominant with
respect to the linear one. We thus choose the limit $k\rightarrow \infty$, so
that the transition probabilities $\tau(1|S)$ are given by
\begin{equation}\label{Eq:tau_s}
 \tau(1|S) = \begin{cases}
  0 & \text{if $S<Q$,}\\
  1 & \text{if $S > R-Q$,}\\
  \dfrac{1}{1+\exp\left\{-2\left[H+Jm(S,R)\right]\right\}} 
    & \text{otherwise.}\\
    \end{cases}
\end{equation}

For $Q=0$ we recover the usual heat-bath 
dynamics for an Ising-like
model with reduced Hamiltonian $\mathcal{H} = \sum_i V(m_i) \sigma_i$.
For $R=3$, $Q=1$ we have the model of Ref.~\cite{Bagnoli_2001} with
two absorbing states. 
The quantities $H$ and $J$ range from $-\infty$ to $\infty$. 
For easy  plotting, we use 
$j = [1+\exp(-2J)]^{-1}$ and $h = [1+\exp(-2H)]^{-1}$ 
as control parameters, mapping the real axis ($-\infty, \infty$) to
the interval $[0,1]$. 

The fraction $c$ of ones in a
configuration and the concentration of clusters $\rho$ are defined by
\[
  c=\dfrac{1}{L}\sum_i s_i\quad\text{and}\quad%
 \rho=\dfrac{1}{L}\sum_i |s_i - s_{i+1}|.   
\]
Both the uniform zero-state and
one-state correspond to $\rho\rightarrow 0$ in the thermodynamic
limit, while the active state corresponds to $\rho > 0$.


\section{Numerical Results}
\label{sec:results}

\begin{figure}
\includegraphics[width=8cm]{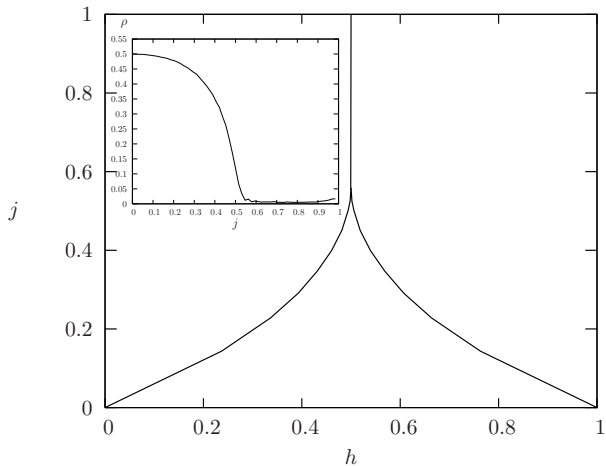}
\caption{\label{figuraR3Q1}
Phase diagram of $R3Q1$ model. The upper--left region
corresponds to the absorbing state 0, the upper--right region to the
absorbing state 1 and the lower region to the disordered phase. In the
inset we show the asymptotic value of $\rho$ as a function of $j$ 
for $H=0$ ($h=0.5$).}
\end{figure}

\begin{figure}
\begin{tabular}{cc}
\includegraphics[width=4cm]{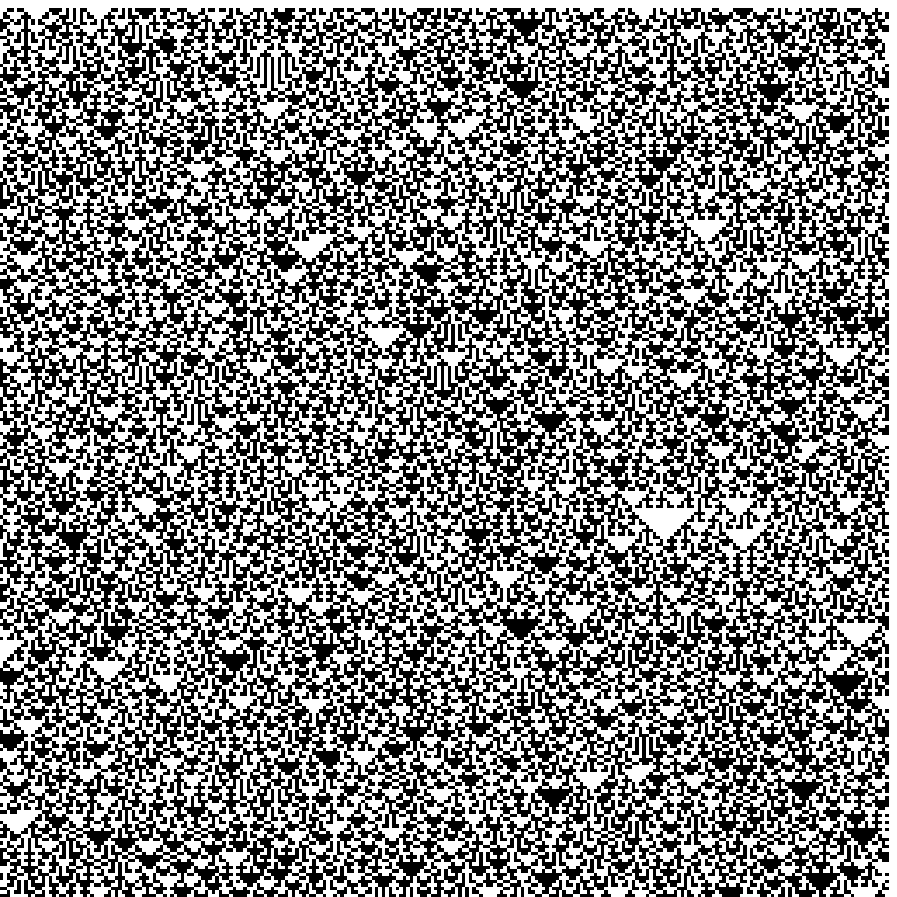} &
\includegraphics[width=4cm]{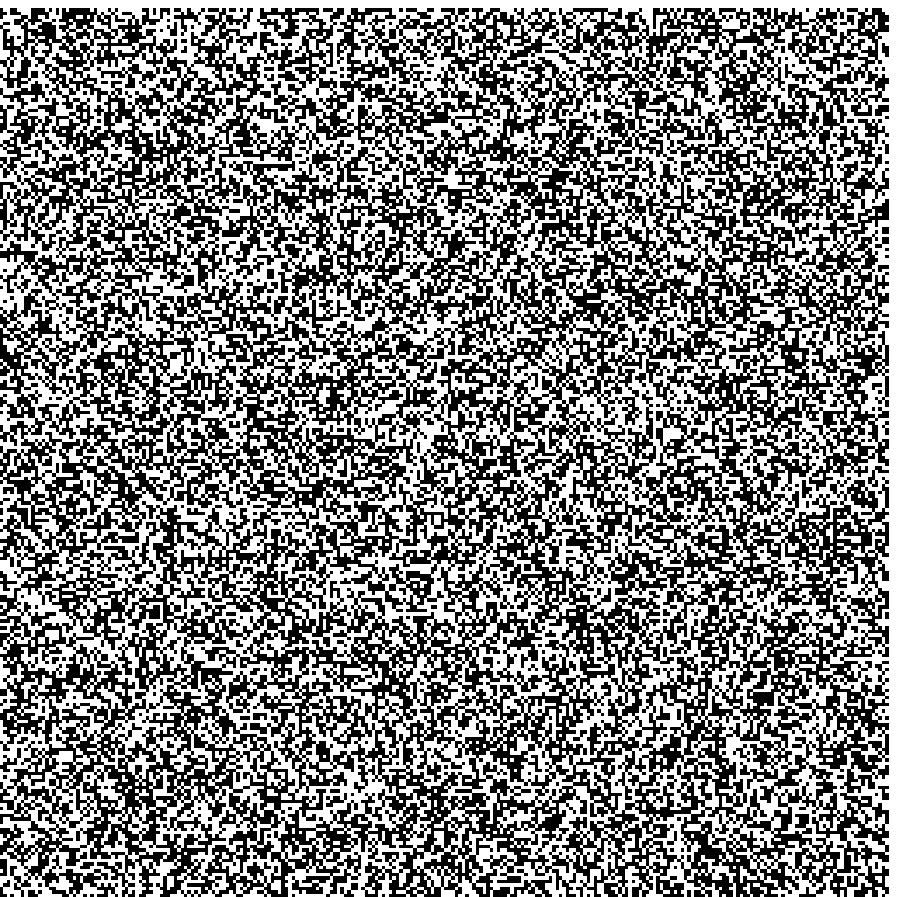} \\
\includegraphics[width=4cm]{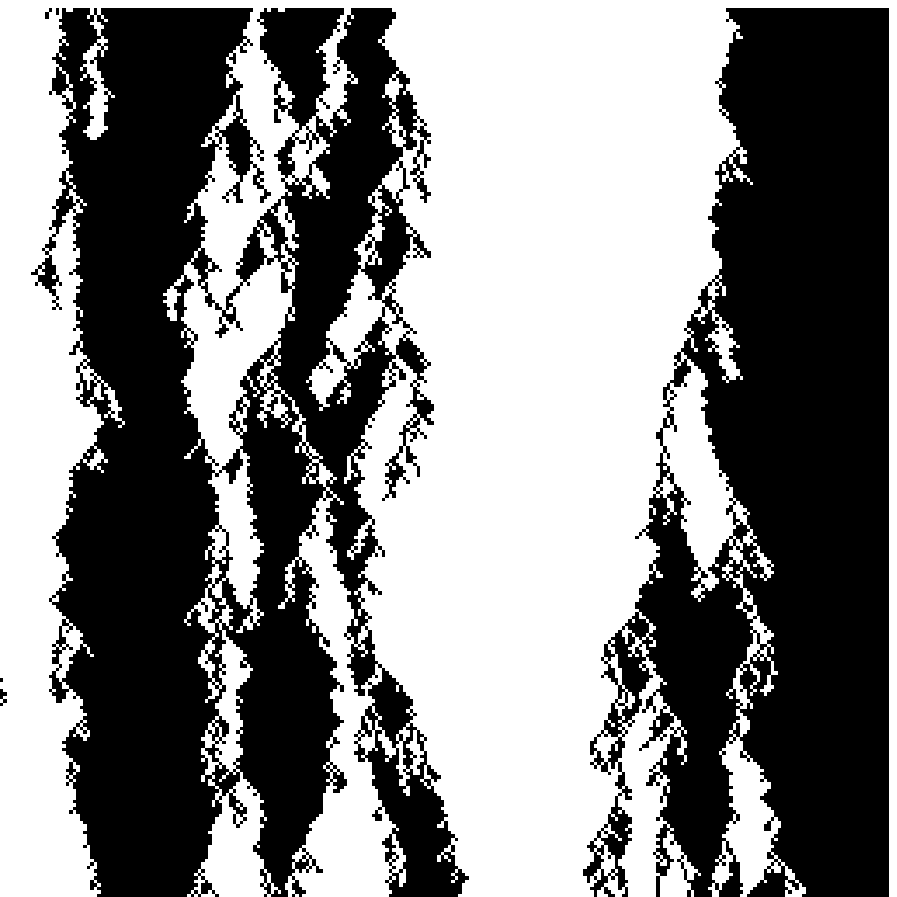} & \\
\end{tabular}
\caption{\label{patternsR3}
Typical space time patterns of the $R3Q1$ model for $H=0$ ($h=0.5$) 
with $L=256$.
(A~--~top left) Chaotic deterministic pattern $j=0.03125$.
(B~--~top right) Disordered phase $j=0.5$.
(C~--~bottom left) competition between quiescent phases $j=0.54688$
(first-order  transition). 
}
\end{figure}

Let us first review the $R3Q1$ case, whose phase diagram is reported in
Fig.~\ref{figuraR3Q1}. A first order phase transition occurs on the vertical
line that ends at a bicritical point where it meets two second order phase
transition curves. In the neighborhood of the first order phase transition line
a typical pattern of the system is composed by large patches of zeroes and
ones, separated by disordered zones (walls) whose width does not grow in time.
These walls perform a sort of random motion and annihilate in pairs (see
Fig.~\ref{patternsR3}-C). By lowering $J$ an active, disordered phase appears
(see Fig.~\ref{patternsR3}-B). The transition between the ordered (absorbing)
and disordered (active) phases occurs by destabilization of the width of the
walls, that percolate in the whole system. In the limit $J=-\infty$ ($j=0$) we
have typical ``chaotic'' class-3 cellular automata
patterns~\cite{wolframclasses}, as shown in
Fig.~\ref{patternsR3}-A. Damage spreading analysis~\cite{Bagnoli_2001}
show that inside the active phase there is a region in which a
variation in the initial configuration can influence the asymptotic
configuration. This region may be called chaotic. However, in the mean
field approximation, this region simply corresponds to a stable fixed
point of the average density. Indeed, the fluctuations of the density
remains quite small and a spatial coarse-graining would generate a simple
random pattern. 

This underlining ``disorder'' is revealed also by 
the asymptotic value of $\rho$ for $H=0$ ($h=0.5$)
as a function of $j$, shown in the inset of Fig.~\ref{figuraR3Q1}. 
It  exhibits a monotonic behavior, with
a maximum for $j=0$ (corresponding to the smallest average length of
homogeneous clusters) and a continuous phase transition at the bicritical
point, that separates the active ($\rho>0$) and quiescent ($\rho=0$)
phases.

\begin{figure}
\begin{tabular}{cc}
\includegraphics[width=4cm]{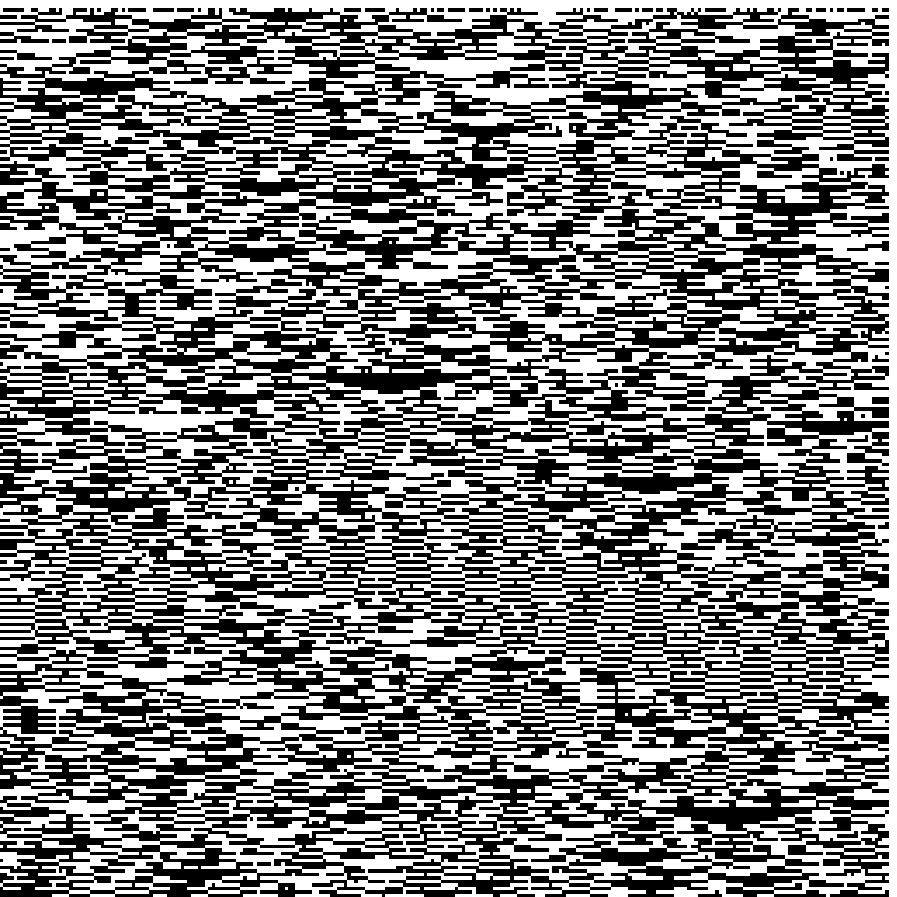} &
\includegraphics[width=4cm]{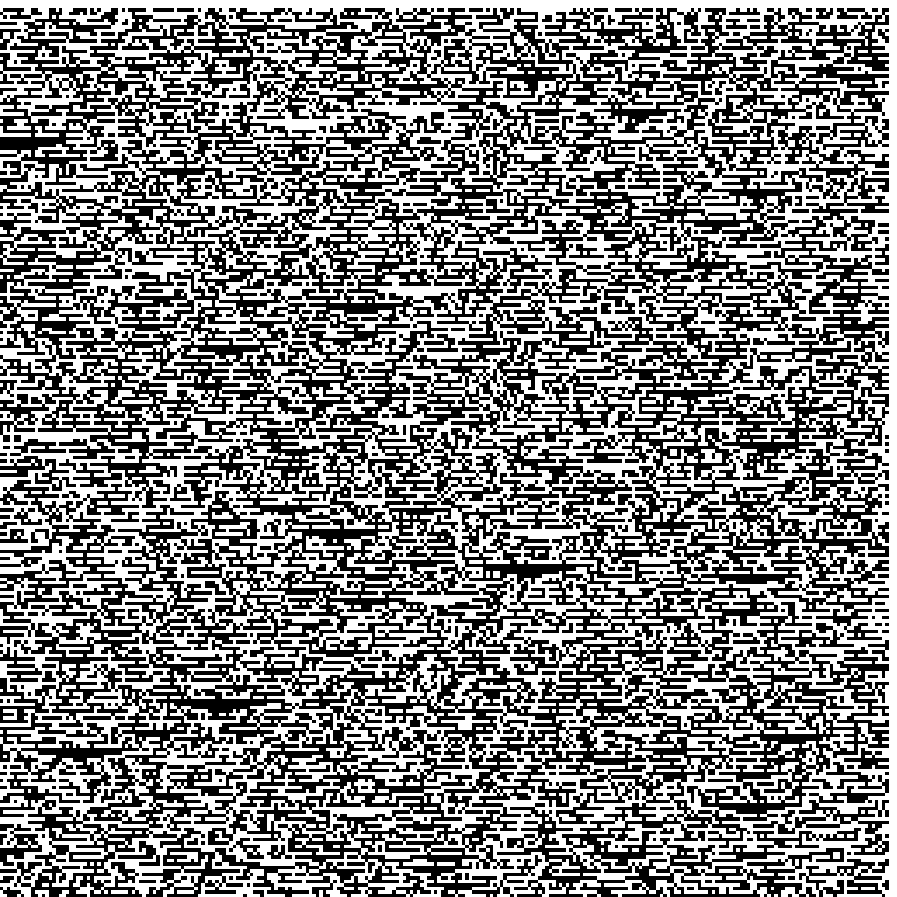} \\
\includegraphics[width=4cm]{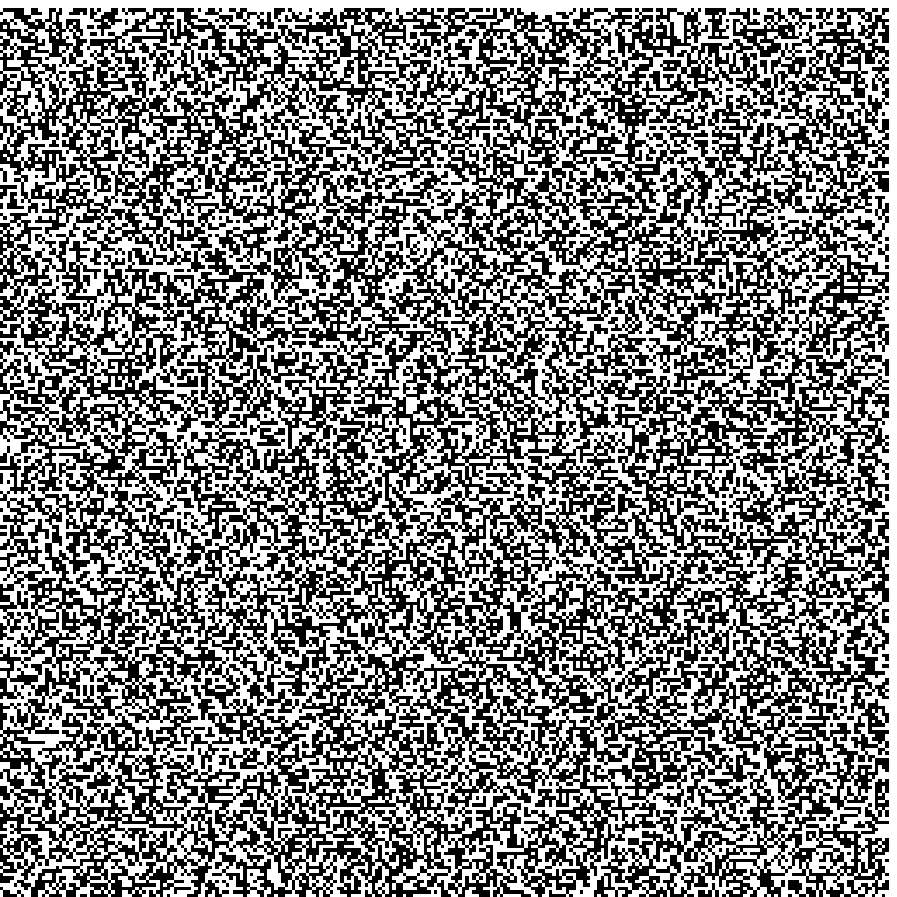} &
\includegraphics[width=4cm]{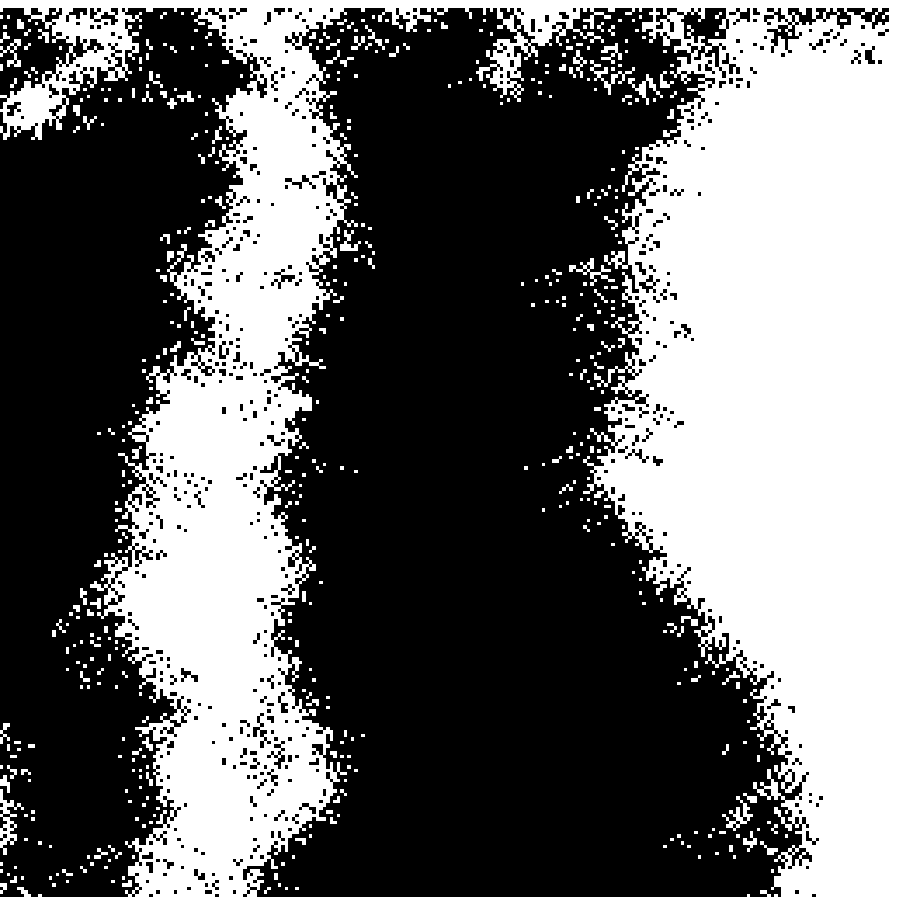}\\
\end{tabular}
\caption{\label{fig:patterns}
Typical space time patterns ($L=256$) for model $R11Q1$ and $H=0$,
starting from a disordered initial condition with $\rho_0=0.5$. 
(A) [top left, $j=0.056250$]: ``coherently-chaotic'' phase. 
One can observe rare 
``triangles'' with ``base'' larger that $R$, 
corresponding to the unstable absorbing states, and other metastable
patterns, corresponding to additional absorbing states for
$J\rightarrow -\infty$.
(B) [top right, $j=0.421875$]: At the boundary between the chaotic and
the irregular phase, 
the only locally absorbing states are (rare) patches of 0 and 1. Most
of the pattern looks random, with some ``triangles'' reminiscent of
chaotic CA. 
(C) [bottom left, $j=0.478125$]: Disordered phase. the pattern looks
random and the difference between patterns obtained with different
values of the parameters (and the same random numbers) is vanishing.
(D) [bottom right, $j=0.562500$]: 
Quiescent phase. In this phase the only stable states are the absorbing
ones. The boundaries separating the phases move randomly until coalescence.}
\end{figure}
\begin{figure}
\includegraphics[width=8cm]{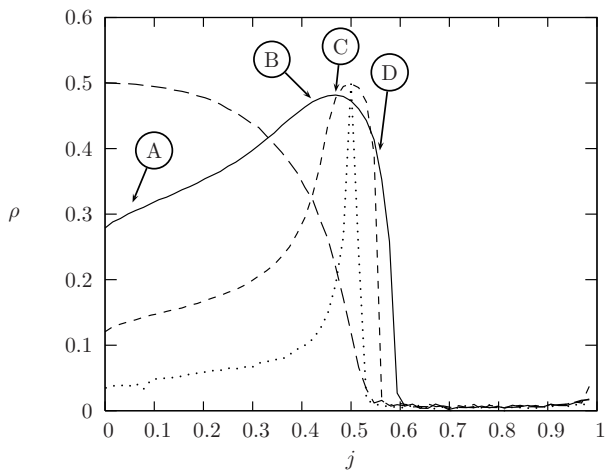} 
\caption{\label{fig:rho}Behavior of $\rho$ for $H=0$. $R3Q1$ (long dashed),
  $R11Q1$ (solid), $R41Q1$ (dashed)
and $R81Q1$ (dotted). Letters correspond to patterns in
Fig~\protect\ref{fig:patterns}.}
\end{figure}
 
By increasing $R$, new features appear.  Typical patterns are
shown in Fig.~\ref{fig:patterns} for different values of $j$. 
As illustrated in Fig.~\ref{fig:rho}, $\rho$ is no longer a monotonic
function  of $j$, and a new, \emph{less disordered} region appears
inside the active one for small values of $j$. The transition between
the active and the quiescent phases become sharper with increasing
$R$, as shown in Fig.~\ref{fig:rho}.

The pattern shown in Fig.~\ref{fig:patterns}-a is reminiscent  of
``chaotic'' deterministic cellular automata. We refer to
this region as \emph{coherently-chaotic},  since it corresponds to
``irregular'' coherent oscillations of large  patches of sites. This
region of the parameters is dominated by an almost-deterministic
behavior, and the presence of many metastable ``absorbing'' states,
revealed by transient regular patterns in  Fig.~\ref{fig:patterns}-a.
 As we discuss in 
Sec.~\ref{sec:approximations},   the mean field analysis gives a
chaotic map for large antiferromagnetic  values of $J$, 
but we were unable to find a clear order parameter that numerically
distinguishes this region from the broader  ``chaotic'' one in
which the asymptotic configuration exhibits a dependence on variations in the
initial  configuration. In Sec.~\ref{sec:lyap}, we analyze the
chaotic region  by means of (finite size) Lyapunov  exponents. 

By increasing $J$, the coherent
patches shrink and the Lyapunov exponent decreases and finally
becomes negative. The system asymptotically loses its dependence on 
the initial
conditions and is dominated by the stochastic components.
This phase is termed \emph{irregular}, and appears completely random. 
Fig.~\ref{fig:patterns}-b shows a typical pattern at the boundary
between the chaotic and the irregular phase. 

Our simulations were carried out using the fragment
method~\cite{fragment}, in which  
a set of configurations (replicas) are
updated with different values of parameters using the same random numbers,
and the same initial configuration. 

By observing in rapid sequence the patterns generated with increasing
values of $J$, one observes a sudden ``freezing'' of the (random)
image, just before the transition between the active to quiescent
phases, corresponding to Fig.~\ref{fig:patterns}-c. This
effect is due to the insensitivity of the patterns not only to the
initial condition, but also to the parameters, see
Sec~\ref{Sec:irregular-disorder}.  We term this phase
\emph{disordered}. 

Finally, the quiescent phase is
asymptotically dominated by the two absorbing states, with the usual
annihilating walls dynamics, Fig.~\ref{fig:patterns}-d.

\begin{figure}
\includegraphics[width=8cm]{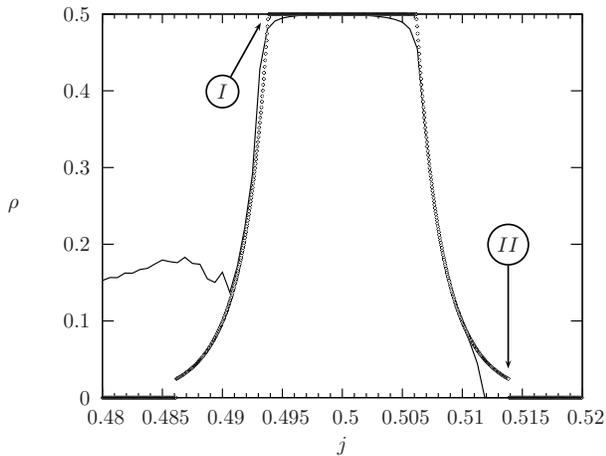}
\caption{\label{fig:H0}
Comparisons between numerical (thin line)  and mean-field (thick, dotted line) 
results for  $R81Q1$ and $H=0$ ($h=0.5$). The estimated critical values are
$j^*_{\text{I}} \simeq 0.493827$ and $j^*_{\text{II}} \simeq 0.51384$.
This figure has been obtained with much larger simulations of the corresponding line in Fig.~\protect\ref{fig:rho}, and more details emerge. The patterns shown in Fig.~\protect\ref{fig:patterns} (which however refer to $R=11$) illustrate the typical behavior of the system to the left (B), in the middle (C) and to the right (D) of the plateau.}
\end{figure}

\begin{figure}
\includegraphics[width=8cm]{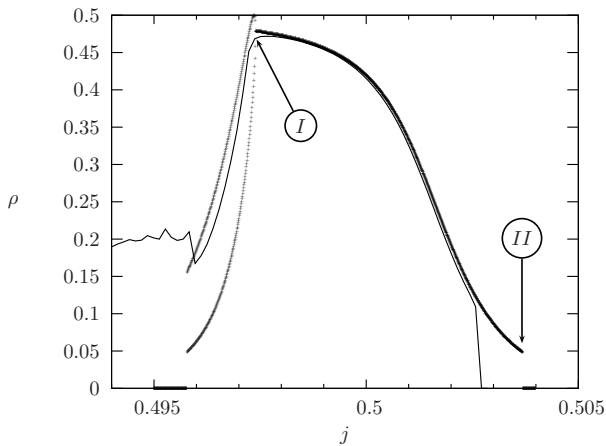}
\caption{\label{fig:H042}
Comparisons between numerical (thin line)  and mean-field (thick, dotted line) 
results for $R201Q5$ and $H=0.42$ ($h=0.7$). The two set of pluses
mark the period-two region.
The estimated critical values are
$j^*_{\text{I}} \simeq 0.49740$ and $j^*_{\text{II}} \simeq 0.50369$.}
\end{figure}

Fig.~\ref{fig:rho} shows that the active-quiescent transition becomes
sharper when $R$ is increased, and that the  cluster density 
exhibits a cusp at the transition.  By enlarging this region one sees
that for $H=0$ (Fig.~\ref{fig:H0}) the cluster density exhibits two 
sharp bends, while for $H =0.42$, (Fig.~\ref{fig:H042}), only one
bend  is present. This is reminiscent of the universality class
change  (parity conservation, DP) for $H=0$, $H \ne 0$, respectively, in the
R3Q1  model~\cite{Bagnoli_2001}. Notice also that the
irregular--disorder transition  at $j \simeq 0.494$ is rather
peculiar, since $\rho$ first decreases and  then suddenly jumps to
high values.

The bends are not finite size or time effects. As shown in
Fig.~\ref{fig:patterns}-c, 
in this range of parameters the probability of observing
a local absorbing configuration (i.e.\ patches of zeroes or ones) is
vanishing. All other local configurations have finite probability of
originating zeroes or ones in the next time step. The observed
transitions are essentially equivalent to those of an equilibrium system,
that in one dimension and for short-range interactions cannot exhibit
a true phase transition. The bends thus become real salient points only
in the limit $R\rightarrow \infty$. 

\subsection{Chaos and Lyapunov exponent in cellular automata}
\label{sec:lyap}

State variables in cellular automata are discrete, and thus the usual linear
stability analysis classifies them as stable systems.  The occurrence of
disordered patterns and their propagation in stable dynamical systems can be
classified into two main groups: \emph{transient chaos} and \emph{stable
chaos}. 

\begin{figure}
\includegraphics[width=8cm]{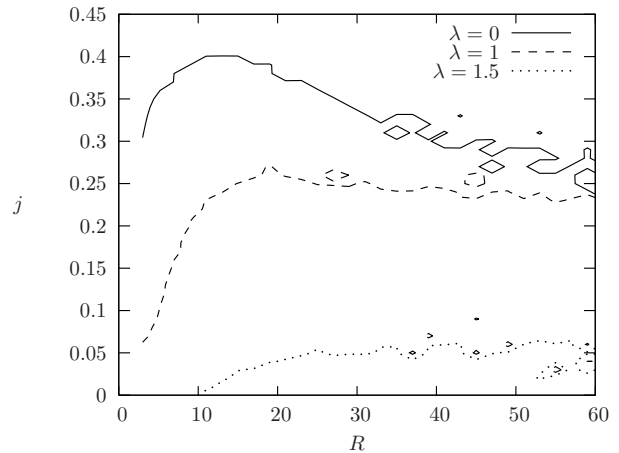}
\caption{\label{fig:Sim}
Contour plot of the maximum Lyapunov $\lambda$ exponents for different values
of $R$ and  $j$, for $H=0$. The solid line represent
the boundary between the $\lambda\ge 0$ phase and the $\lambda=-\infty$ one.}
\end{figure}

Transient chaos  is an irregular behavior of finite lifetime characterized by
the coexistence in the phase space of stable attractors and chaotic non
attracting sets, namely chaotic saddles or repellers~\cite{Tel}. After a
transient irregular behavior, the system generally collapses abruptly onto a
non-chaotic attractor. This kind of behavior is reminiscent of some 
{\em class-4}
(nonchaotic) deterministic cellular automata
(DCA)~\cite{wolframclasses}, and can
be present in continuous systems having a discrete dynamics as a limiting
case~\cite{stablechaos}.

Stable chaos constitutes a different kind of transient irregular
behavior \cite{CK88,PLOK} which cannot be ascribed to the presence of
chaotic saddles and therefore to divergence of nearby trajectories.
Moreover, the lifetime of this transient regime may scale
exponentially with the system size
(supertransients as defined in Refs.~\cite{CK88,PLOK}), 
and the final stable attractor is
practically never reached for large enough systems. One is thus
allowed to assume that such transients may be of substantial 
experimental interest and become the only physically relevant states
in the  thermodynamic limit. 

The emergence of this ``chaoticity'' in {\em class-3} (chaotic) 
DCA dynamics, is effectively illustrated by the damage spreading
analysis~\cite{Damage1,Damage2}, which measures the sensitivity to initial
conditions and for this reason is considered as the natural extension of the
Lyapunov technique to discrete systems. In this method, indeed, one monitors
the behavior of the distance between  two replicas of the system evolving from
slightly different initial conditions, or the speed of propagation of a
disturbance~\cite{wolfram}. The dynamics is considered unstable and
the DCA is called chaotic, whenever a small initial difference between replicas
spreads through the whole system.  On the contrary, if the initial difference
eventually freezes or disappears,  the DCA is considered non chaotic.  

Due to the limited number of states of the automaton, damage spreading
does not account for the maximal ``production of uncertainty'', since the two
replicas may synchronize locally just by chance (self-annihilation of the
damage). Moreover, there are different definitions of damage spreading for the
same rule~\cite{DomanyHinrichsen}.

To better understand the nature of the active phase, and up to what extent it
can be denoted \emph{chaotic},  we extend the finite-distance Lyapunov exponent
definition~\cite{LyapunovCA} to probabilistic cellular automata. A similar
approach has been used in Ref.~\cite{luque}, calculating the Lyapunov exponents
of a Kauffman random Boolean network in the annealed approximation. As shown in
this latter paper, this computation gives the value of the (classic) Lyapunov
exponent obtained by the analysis of time-series data using the Wolf
algorithm~\cite{wolf}.

Given a Boolean function $f(x,y,\dots)$, 
we define the Boolean derivative $\de f/\de x$ with respect to $x$ by
\begin{equation}
 \label{eq:boolean}
 \dfrac{\de f}{\de x} = \begin{cases} 
  1 & \text{if $f(|x-1|, y, \dots)  \neq f(x,y,\dots)$,}\\
  0 & \text{otherwise,}
 \end{cases}
\end{equation}
which represents a measure of sensitivity of a function with respect to $x$.
The evolution rule of a probabilistic cellular automaton 
can be written as
\[
 x_i^{t+1}=\begin{cases}
           1 & r<\tau(1|S_i^t),\\
		   0 & \text{otherwise}
           \end{cases}
\]
where $r$ is a random number uniformly distributed in $[0,1]$. The Boolean
derivative with respect to $x_j$ is evaluated by using the same $r$
in the comparison implied in Eq.~(\ref{eq:boolean}).

For a PCA, we can thus build the Jacobian matrix $J_{ij} = \de
x_i^{t+1}/\de x_j^t$, $i,j=0,\dots,L-1$.  
The maximum Lyapunov exponent $\lambda$ can now be defined
in the usual way as the expansion rate of a tangent 
vector $\v{v}(t)$, whose time evolution  is given by
\[
 \v{v}(t+1) = \v{J} \v{v}(t).
\]
When all the components of $\v{v}$ become zero,
$\lambda=-\infty$ and no information about the initial configuration may
``percolate'' as $t\to\infty$~\cite{LyapunovCA}. This maximum Lyapunov 
exponent is also related to the synchronization properties of 
cellular automata~\cite{synca}. 

A preliminary numerical computation of $\lambda$ for our model is reported in
Fig.~\ref{fig:Sim}. It can be noticed that the boundary $j_c(R)$ of
$\lambda\ge0$ is not monotonic with $R$, reaching a maximum value for $R\simeq
11$. By comparisons with Fig.~\ref{fig:rho}, one can see that the chaotic phase
is included in the irregular one.

\begin{figure}
\includegraphics[width=8cm]{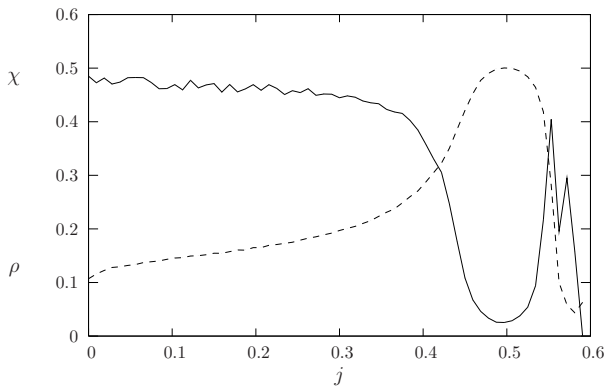}
\caption{\label{fig:diff}
Susceptibility $\chi$ (continuous line) and average
number of clusters $\rho$ (dashed line) for $R11Q1$, $H=0$,
$\Delta j = 0.01$}

\end{figure}

\subsection{Parameter dependence}
\label{Sec:irregular-disorder}

We can numerically  characterize the irregular-disorder  transition  by looking at the
sensitivity of space-time patterns with respect to the variations of parameters.
Using the fragment method~\cite{fragment} a set of configurations (replicas) are
updated with different values of parameters using the same random numbers,
i.e.\ with the same disordered field and initial conditions. 
The quantity
\[
\chi(J) = \lim_{\Delta j \rightarrow 0} \lim_{t\rightarrow
 \infty}  \sum_i\left| s_i^t(J+\Delta J) - s_i^t(J)\right|  
\]
is a measure of the susceptibility. For large correlations one expects
very small differences among replicas. As shown in
Fig.~\ref{fig:diff}, this susceptibility is strongly correlated to
the average number of clusters $\rho$. 
%


\section{Mean field approximation}
\label{sec:approximations}

\begin{figure}
\includegraphics[width=8cm]{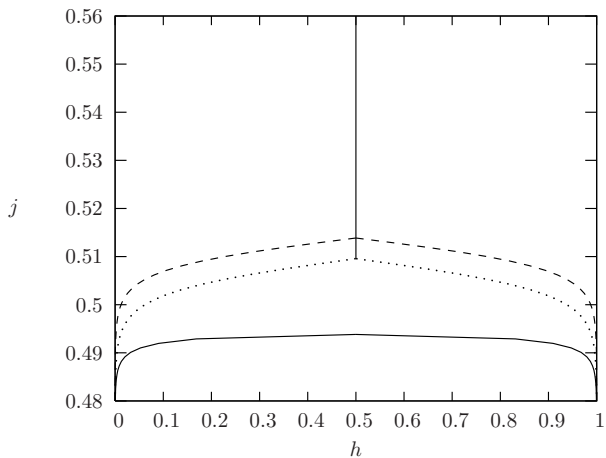}
\caption{\label{fig:meanfieldphase}
Mean-field phase diagram  for $R=81$. The upper curves correspond to the
transition II from the quiescent to the disordered phases, $R81Q5$ (dashed)
and $R81Q1$ (dotted). The solid curve corresponds to the transition I
from the disordered to the irregular phase. 
}
\end{figure}

In the simplest mean field approximation, the
density $c$ evolves in time according to
\begin{equation}\label{rmf}
  c(t+1) =  \sum_{S=0}^{R} W^{(R,S)}(c(t))\tau(1|S),
\end{equation}
with
\[
 W^{(R,S)}(c) = \binom{R}{S} c^S (1-c)^{R-S}.
\]
An example of a mean-field phase diagram is reported in
Fig.~\ref{fig:meanfieldphase}.

For large $R$, $W^{(R,S)}$ can be approximated by
\begin{equation}\label{gaussian}
  W^{(R,S)}(c) \simeq \dfrac{1}{\sqrt{2\pi
  c(1-c)R}}\exp\left[-\dfrac{R(S/R -c)^2}{2c(1-c)}\right],
\end{equation}
and the summation can be replaced by an integral.
The parameters of the resulting  equation may be rescaled 
by using $\tilde J = JR$ and $\tilde Q=Q/R$. 
In the limit $R \rightarrow \infty$ ($\tilde J$ and $\tilde Q$ fixed),
$c(t+1) =  f(c(t); H, \tilde{J}) $
with 
\begin{equation}\label{mf}
  f = \begin{cases}
  0 & \text{if $c<\tilde{Q}$,}\\
  1 & \text{if $c > 1-\tilde{Q}$,}\\
  \dfrac{1}{1+\exp(-2(H+\tilde{J}(2c-1))} & \text{otherwise.}\\
    \end{cases}
\end{equation}
The requirement of constant rescaled variables means that
the variations of $J$ for large $R$ triggers large variations of
$\tilde J$, and consequently the transition region becomes sharper.

This approximation is supposed to be particularly good in the
disordered phase, i.e.\ for $J, H\simeq 0$, as shown in 
Fig.~\ref{fig:H0}. In the finite-dimensional
case, one has to add a noise term, representing the influence of
neighboring sites, whose amplitude is of the order of the width of
the Gaussian in Eq.~(\ref{gaussian}),  $\sqrt{c(1-c)/R}$. 

In this mean-field approximation, the order parameters $c$ and the
numerical density of clusters $\rho$
are related by $\rho=2c(1-c)$. 

For $J, H\simeq 0$ the mean field approximation gives a stable fixed
point $c^*(H, \tilde J) \ne 0, 1$ for the density, which assumes the
maximum value $c^* =0.5$ for $J=H=0$. 
 
\begin{figure}
\includegraphics[width=8cm]{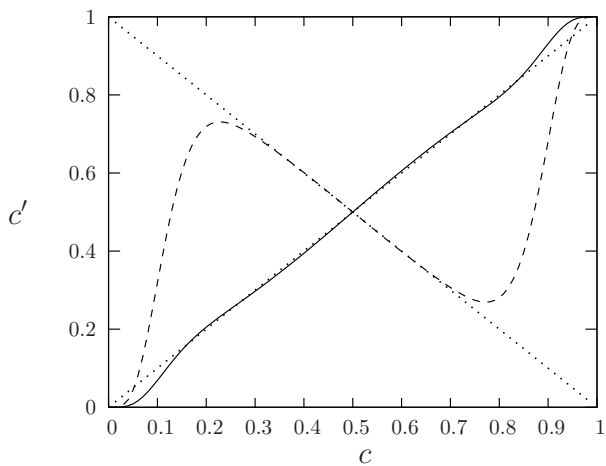}
\caption{ \label{fig:meanfield}
Mean field map
Eq.~(\ref{rmf}) in the neighborhood of transition I ($\tilde
J\simeq -1$, dashed line) and II
($\tilde
J\simeq 1$, solid line),  for $H=0$ and $\tilde Q=0.02$}
\end{figure}

The origin of the irregular-disordered and 
disorder--quiescent phase transitions marked by (I) and (II) respectively in
Figs.~\ref{fig:H0} and \ref{fig:H042}
is due to the loss of stability of the mean-field fixed point $c^* \ne 0, 1$
(see Eq.~(\ref{mf}) and Fig.~\ref{fig:meanfield}).

The transition I is given by
\begin{equation}\label{transI}
  c^*_{\text{I}} = f(c^*_{\text{I}}; H, \tilde{J}); \qquad
  \left.\dfrac{ \text{d} f(c; H, \tilde{J})}{\text{d}
  c}\right|_{\text{I}} = -1.
\end{equation}
which would induce a period-doubling cascade in the mean-field
approximation (disregarding the presence of absorbing states). 
The solution of Eq.~(\ref{transI}) 
is
\begin{align*}
 c^*_{\text{I}}&= \dfrac{1}{2}\left(1+\sqrt{1+\tilde{J}^{-1}}\right),\\
 H^*_{\text{I}}& = - \tilde{J}\sqrt{1+\tilde{J}^{-1}} - 
 \dfrac{1}{2}\log\left(\dfrac{1-\sqrt{1+\tilde{J}^{-1}}}
 {1+\sqrt{1+\tilde{J}^{-1}}}\right).
\end{align*}
The critical value $J^*_{\text{I}}$ can be
obtained numerically once given the value of $H^*=H^*(J^*)$, and
corresponds to the left bend in Fig.~\ref{fig:H0} and
\ref{fig:H042}.

For $H=0$ (Fig.~\ref{fig:H0}) the period-doubling instability
brings the local
configuration into an absorbing state, and the lattice dynamics is
therefore driven by interactions among patches which are locally 
absorbing. This essentially corresponds to the dynamics of a
Deterministic Cellular Automata (DCA) of \emph{chaotic} type, i.e.\ 
a system which is insensitive to infinitesimal perturbations but
reacts in an unpredictable way to \emph{finite} perturbations
\cite{stablechaos}. In other words, in this region the original stochastic
model behaves like a ``chaotic'' deterministic one after a coarse-graining of
patches.  We expect that this correspondence will become more and more exact
with growing $R$. From a theoretic-field point of view this means that the
renormalization flux tends towards a ``chaotic'' model instead of the usual
fixed-point dynamics. 

For $H=0.42$, as shown in Fig.~\ref{fig:H042}, the period-two phase has a
finite amplitude, before falling into the DCA-like dynamics by reducing $J$.

For $J>0$ (transition II) we have
\[
  c^*_{\text{II}} = f(c^*_{\text{II}}; H, \tilde{J}) = 
    \begin{cases}\tilde Q \\ 1-\tilde
  Q\\\end{cases}
\]
and thus the critical value $J^*_{\text{II}}$ is 
\begin{equation*}
  J^*_{\text{II}} = \dfrac{1}{2\tilde Q -1} \left[\pm H + \dfrac{1}{2}
  \log \left(\dfrac{\tilde Q}{1-\tilde Q}\right)\right].
\end{equation*}
Once that a portion of the lattice has been
attracted to an absorbing state, it pulls the neighboring regions to
this same state, due to the  ferromagnetic ($J>0$) coupling. 

This approximation, disregarding fluctuations, overestimates the
critical value $J^*_{\text{II}}$, as shown in Figs.~\ref{fig:H0} and
\ref{fig:H042}.

\begin{figure}[t]
\includegraphics[width=6cm]{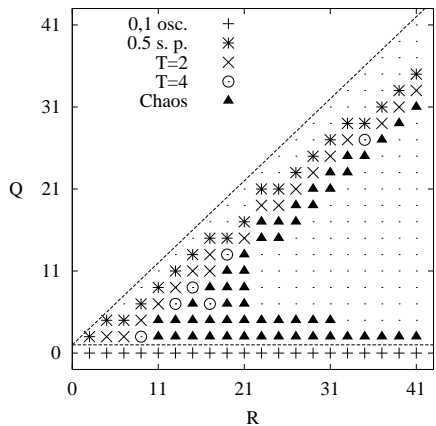}
\caption{\label{fasecm} Case $H=0$, $J=-\infty$: the
mean-field $R-Q$ phase diagram of the mean-field approximation. 
Points mark parameter values for which the
absorbing states are  the only stable attractors. 
A plus sign denotes period-2 temporal 
oscillations between absorbing states, 
a star denotes the presence of a stable
point at $c=0.5$, a cross (circle) denotes period-two (four)
oscillations between two non-zero and non-one densities, triangles
denote chaotic oscillations.} 
\end{figure}

We analyze extensively the mean-field behavior of the system for $H=0$
and $J=-\infty$. As shown in 
Fig.~\ref{fasecm}, for a given value of $R$,  
there is always a critical $Q_c$ value of $Q$ for
which the active phase disappears, with an approximate correspondence
$Q_c \simeq 2/5 R$. One can also observe that the chaotic
oscillations of the mean-field map, that roughly  corresponds to the
coherent-chaotic phase of the model, appear only for large values of
$R$. Since the absorbing states are always present, there chaotic
oscillations may bring the system into the quiescent phase, as shown
by the ``hole'' to the right of triangles in Fig.~\ref{fasecm}. 

The correspondence between the chaotic behavior of the mean-field
approximation and the actual behavior of the system will be the subject
of a future work.


\section{Conclusions}
\label{sec:conclusions}

We have investigated a general one dimensional model with
extended-range interactions and symmetric absorbing states. The model
is characterized by a competing ferromagnetic linear coupling and an
antiferromagnetic nonlinear one. 

By means of
numerical simulations and mean field approximations we have shown
that a chaotic phase is present
for strong antiferromagnetic coupling. 
This phase may be identified as a stable-chaotic region, in which the
behavior of the (originally stochastic) system is essentially deterministic  
and its behavior is highly irregular and essentially unpredictable. 
The mean field map exhibit a chaotic behavior for a large interaction
range $R$, and this behavior is reflected in the appearance of many
metastable states in the system, for extremely strong
antiferromagnetic coupling. 

A disordered phase, insensitive of parameter variations,
appears at the boundary between  the active and the quiescent ones,  and the
transitions appear to be of equilibrium type, i.e.\ truly
salient points only in the limit $R\rightarrow\infty$.

We expect that in higher dimensions one can recover some aspects of
these  phase transitions  without imposing the presence of absorbing
states, i.e.  using finite couplings. 
 

\section*{Acknowledgements}
Partial economic support from project IN109602 DGAPA--UNAM and the
Coordinaci\'on de la Investigaci\'on Cient\'\i fica UNAM is
acknowledged.


\appendix\section{Equivalence 
between dynamic Ising Model and Cellular Automata}

As an illustration,
we present here a derivation of the equivalence of the kinetic Ising model 
(here we choose heat bath) with a cellular 
automaton (the Domany-Kinzel model) in order to elucidate the  role of 
infinite coupling parameters and absorbing states. This derivation is 
similar to that of Ref.~\cite{Georges} but more general.

The Ising model is defined by the couplings among spins. The configuration at
time $t$ is denoted as $\boldsymbol{\sigma}= \sigma_1, \sigma_2,\dots$ and the
configuration at time $t+1$ as $\boldsymbol{\sigma}'= \sigma'_1,
\sigma'_2,\dots$.
Let us write the temperature-dependent Hamiltonian as 
\[
 \mathcal{H}(\boldsymbol{\sigma}) = \sum_i H(\sigma_{i-1},\sigma_{i}
 \sigma_{i+1}),
\]
with 
\begin{equation}\label{H}
  H(x,w,y) = J^{(0)} w + J^{(1)} x w + J^{(2)} w y + J^{(3)} xwy.
\end{equation}

The transition probabilities $\tau$ must obey the detailed balance condition
\[
  \dfrac{\tau(\boldsymbol{\sigma}'|\boldsymbol{\sigma})}
    {\tau(\boldsymbol{\sigma}|\boldsymbol{\sigma}')} = 
      \exp\bigl(\mathcal{H}(\boldsymbol{\sigma}')-\mathcal{H}(\boldsymbol{\sigma})\bigr).
\]
In each step we can update in parallel all even or odd sites, obtaining 
\[
 \dfrac{\tau(\boldsymbol{\sigma}'|\boldsymbol{\sigma})}
    {\tau(\boldsymbol{\sigma}|\boldsymbol{\sigma}')} =   
  \prod_{i}'
  \dfrac{\tau(\sigma_{i-1},\sigma'_i, \sigma_{i+1}|\sigma_{i-1},\sigma_i,
  \sigma_{i+1})}
    {\tau(\sigma_{i-1},\sigma_i ,\sigma_{i+1}|\sigma_{i-1},\sigma'_i,
  \sigma_{i+1})} ,
\] 
where the product is restricted to either even or odd sites.  
The detailed balance condition can thus be satisfied locally. 
Choosing the Heat Bath dynamics
\[
  \tau(x,1,y|x,w,y) = \dfrac{\exp\bigl(-H(x,1,y)\bigr)}
  {\exp\bigl(-H(x,-1,y)\bigr)+\exp\bigl(-H(x,1,y)\bigr)}
\]
the
transition probabilities does not depend on the present value of the
spin $w=\sigma_i^t$ so that the lattice (with even or infinite lattice
sites) may be decoupled into two noninteracting sublattices with the
same geometry of the DK model. From now on, 
to be coherent with the usual cellular
automaton notation, we shall express the transition probabilities in
terms of the Boolean variables $s_i=(\sigma_i+1)/2$,  
and we shall denote the local field 
as $h(a,b)\equiv H(2a-1,1,2b-1)$.
 
Let us denote the DK transition probabilities $\tau(s'_i|s_{i-1},
s_{i+1})$ as
\[
  \tau(1|00)=\varepsilon, \qquad \tau(1|01)=\tau(1|10)=p, \qquad \tau(1|11)=q.
\]

We get for the heat bath dynamics
\[
  \tau(1|ab) = \dfrac{1}{1+\exp\bigl(-2h(a,b)\bigr)}
\]
and thus
\[
  h(a,b) = \dfrac{1}{2}\ln \dfrac{\tau(1|ab)}{1-\tau(1|ab)}.
\]

Substituting into Eq.~\eqref{H} one obtains a linear system
\[
  \begin{split}
    J^{(0)} -  J^{(1)} -  J^{(2)} +  J^{(3)} &= \frac{1}{2}\ln
      \frac{\varepsilon}{1-\varepsilon},\\
    J^{(0)} +  J^{(1)} -  J^{(2)} -  J^{(3)} &= \frac{1}{2}\ln \frac{p}{1-p},\\
    J^{(0)} -  J^{(1)} +  J^{(2)} -  J^{(3)} &= \frac{1}{2}\ln \frac{p}{1-p},\\
    J^{(0)} +  J^{(1)} +  J^{(2)} +  J^{(3)} &= \frac{1}{2}\ln \frac{q}{1-q}.
  \end{split}
\]
\vspace{3cm}

Finally, we have
\[
  \begin{split}
  J^{(0)}  &= \frac{1}{8}\ln  \frac{\varepsilon}{1-\varepsilon} +
    \frac{1}{8}\ln \frac{q}{1-q} + \frac{1}{4}\ln \frac{p}{1-p}, \\
  J^{(1)} =J^{(2)}  &= - \frac{1}{8}\ln  \frac{\varepsilon}{1-\varepsilon} +
    \frac{1}{8}\ln \frac{q}{1-q} , \\
  J^{(3)}   &= \frac{1}{8}\ln  \frac{\varepsilon}{1-\varepsilon} +
    \frac{1}{8}\ln \frac{q}{1-q} - \frac{1}{4}\ln \frac{p}{1-p}. \\    
  \end{split}
\]

In the limit $\varepsilon \rightarrow 0$ all couplings become infinite.

\end{document}